\documentclass[10pt,twocolumn,a4paper]{article}
\usepackage[american]{babel}
\usepackage[utf8]{inputenc}
\usepackage{subcaption}
\usepackage{graphicx}
\usepackage[export]{adjustbox}
\usepackage{amsmath}
\usepackage{amssymb}
\usepackage{tabularx}
\usepackage{multirow}
\usepackage[table]{xcolor}
\usepackage{float}
\usepackage{flushend}
\usepackage{titlesec}
\titleformat{\section}
  {\normalfont\fontsize{12}{15}\bfseries}{\thesection}{1em}{}
\usepackage{titlesec}
\usepackage{authblk}
\usepackage{lipsum}
\usepackage{cite}
\usepackage{cleveref}
\usepackage{layout}
\usepackage{ulem}
\begin{document}
\title{Conditional Normalizing flow for Monte Carlo sampling in lattice scalar field theory}

\author[1]{Ankur Singha}
\author[1]{Dipankar Chakrabarti}
\author[2]{Vipul Arora}
\affil[1]{Department of Physics, Indian Institute of Technology Kanpur, Kanpur-208016, India}
\affil[2]{Department of Electrical Engineering, Indian Institute of Technology Kanpur, Kanpur-208016, India}

\maketitle
\begin{abstract}

The cost of Monte Carlo sampling of lattice configurations is very high in the critical region of lattice field theory due to the high correlation between the samples. This paper suggests a Conditional Normalizing Flow (C-NF) model for sampling lattice configurations in the critical region to solve the problem of critical slowing down. We train the C-NF model using samples generated by Hybrid Monte Carlo (HMC) in non-critical regions with low simulation costs. The trained C-NF model is employed in the critical region to build a Markov chain of lattice samples with negligible autocorrelation. The C-NF model is used for both interpolation and extrapolation to the critical region of lattice theory. Our proposed method is assessed using the 1+1-dimensional scalar $\phi^4$ theory. This approach enables the construction of lattice ensembles for many parameter values in the critical region, which reduces simulation costs by avoiding the critical slowing down.
\end{abstract}

\section{Introduction}
In lattice field theory, Monte Carlo Simulation techniques are used to sample lattice configurations based on a distribution defined by the action of the lattice theory.
The parameter value at which we generate the lattice samples determines the cost of the simulation. The non-critical region of the lattice theory has low simulation costs for algorithms like Hybrid Monte Carlo (HMC))\cite{duane1987hybrid}. 
However, as we attempt to sample uncorrelated lattice configurations from the critical region, the simulation cost increases rapidly. In the critical region, the integrated autocorrelation time, which gives the measure of correlation, increases rapidly and diverges at the critical point. For a finite-size lattice critical point corresponds to the peak point of the autocorrelation curve. This problem is known as the critical slowing down\cite{Wolff:1989wq,Schaefer:2010hu}. Many efforts have been made to lessen the impact of critical slowing in statistical systems and lattice QFT \cite{Ramos:2012bb, gambhir2015improved,Endres_2015}. But it always remains a challenging task to simulate near the critical point of a lattice QFT by overcoming the critical slowing down.

These days, ML-based solutions to this problem are becoming popular. Various ML algorithms have been applied for statistical physics and condensed matter problems\cite{PhysRevD.100.011501, PhysRevLett.122.080602,Pawlowski_2020,Nicoli_2020, Carrasquilla_2020, Liu_2017, Vielhaben_2021, Chen_2018,PhysRevB.94.165134,Science.355.602,PhysRevB.94.195105,PhysRevLett.120.066401,Singh_2021} .
Some generative learning algorithms\cite{PhysRevD.100.034515,albergoflowbased,shanahan2018machine,Kanwar_2020,albergo2021introduction,Albergo:2022qfi,Hackett:2021idh} have recently been developed to avoid the difficulty in lattice field theory. Conditional GAN 
was used in a 2D lattice Gross Neveue model \cite{Singha:2021nht} and shown to be effective in mitigating the influence of critical slowdown.
However, explicit probability density estimation is not accessible in GAN; thus, we cannot guarantee that the model distribution is identical to the true lattice distribution. 
Flow based generative leanings  are found successful in avoiding the problem of critical slowing down in scalar field theory\cite{PhysRevD.100.034515}, fermionic system\cite{albergoflowbased}, U(1) gauge theory\cite{Kanwar_2020} and Schwinger models\cite{Albergo:2022qfi}. In the flow-based approach \cite{PhysRevD.100.034515,Kanwar_2020} an NF model is trained at a single value of the action parameter with reverse KL divergence. Finally, the trained model can generate lattice samples at the same parameter value. This model is initialized from scratch for a parameter value and does not use any lattice samples. 
Since HMC simulation in the non-critical region is not affected by the critical slowing down problem, we can use samples from that region to train a generative model. 
We present a method for sampling lattice configurations near the critical regions using Conditional Normalizing Flow (C-NF) to reduce the problem of critical slowing down. This method involves training a C-NF model in a non-critical region to produce samples for several parameter values in the critical region. Our goal is to generate lattice configurations from the distribution $p(\phi|\lambda_{crit}) =\frac{1}{Z} e^{-S(\phi|\lambda_{crit})}$, where $\phi $ denotes the lattice field, $\lambda_{crit}$ denotes the action parameter close to critical point and Z is the partition function. 
 We train a C-NF model $\tilde{p}({\phi|\lambda})$ with HMC samples from $p(\phi|\lambda)$ for various non-critical $\lambda$ values.
 We train the C-NF model to be a generalized  model over $\lambda$ parameters. The model is then interpolated or extrapolated to the $\lambda $ values in the critical region to generate lattice configurations. However, the interpolated or extrapolated model may not directly provide samples from the true distribution. But the exactness can be guaranteed by using the Metropolis-Hastings(MH) algorithm at the end. So, after training, we use the interpolated/extrapolated model $\tilde{p}(\phi|{\lambda_{crit}})$ at critical region as proposal for constructing a Markov Chain via an independent MH algorithm\cite{PhysRevD.100.034515}. This method is useful when the probability distribution is known up to a normalizing factor. 

The primary contributions of this study are as follows:
\begin{enumerate}
\item Using Conditional Normalizing Flows, we present a new method for sampling lattice configurations near the critical regions. This method eliminates the critical slowdown problem. 

\item The model has the ability to learn about the lattice system across multiple $\lambda$ values and use this knowledge to generate sample at any given $\lambda $ values. As a result, our model can generate samples at multiple $\lambda$ values values in the critical region, which is not possible for the existing flow-based methods in lattice thoery.

\item We also demonstrate that Conditional Normalizing flow can be used to do both interpolation and extrapolation for lattice $\phi^4$ theory. The extrapolation demonstrates the possibility of using our approach for sampling lattice gauge theory. 
\end{enumerate}

\section{Lattice scalar $\phi^4$ Theory}
In $2d$ euclidean space, the action for $\phi^4$ theory can be written as: 
\begin{align}
S(\phi,\lambda,m)=\int dx^2[(\partial_\mu \phi(x))^2+m^2 \phi(x)^2+\lambda\phi(x)^4]
\end{align}
 where $\lambda $ and $m$ are the two parameters of the theory.
 
 On lattice the action become:
 \begin{align}
 S(\phi,\lambda,m)=&\sum\limits_{x} \Big[ \sum\limits_{\mu={1,2}}
 \Big[2\phi(x)^2+\phi(x)\phi(x+\hat{\mu})\notag\\
 &-\phi(x)\phi(x-\hat{\mu})\Big] +m^2\phi(x)^2 +\lambda\phi(x)^4\Big]
 \end{align}
  
 where $x$ is a $2d$ discrete vector and $\hat{\mu}$ represents two possible directions on the lattice. $\phi(x)$ is defined on each lattice site, taking only real values.
 
\begin{figure}[H]
\centering
      \includegraphics[width=.5\textwidth]{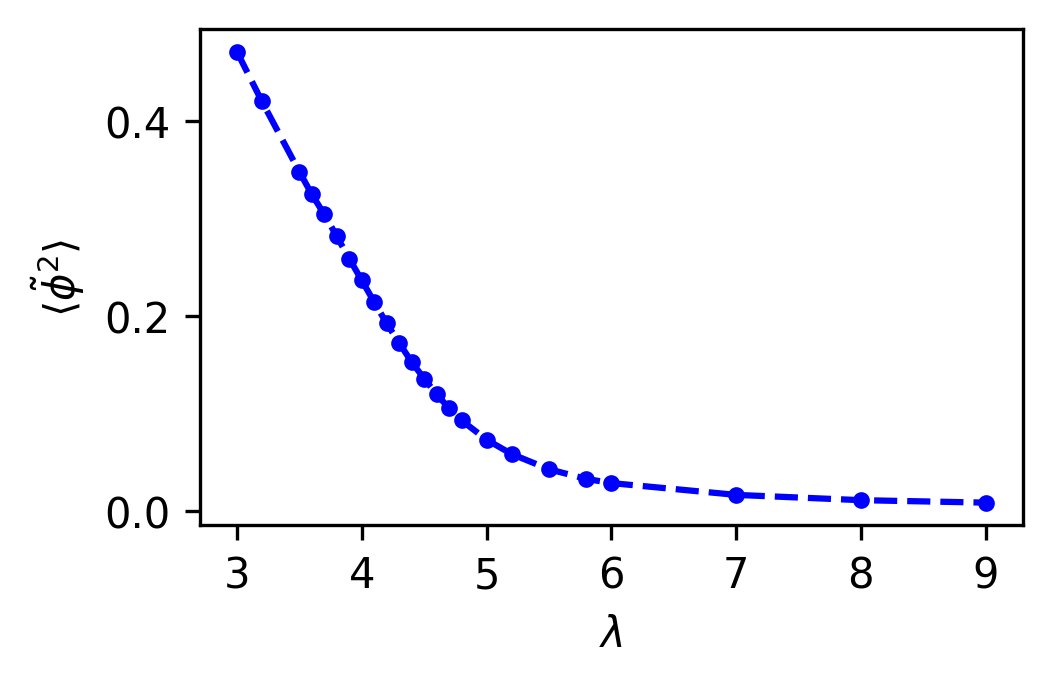}
      \caption{ HMC simulation: order parameter $\langle \Tilde{\phi}^2 \rangle$ vs $\lambda$ for $ m^2=-4$, which shows a  2nd order phase transition.}
      \label{SSB}
\end{figure}

We choose this specific form of action for HMC simulation because it is suited for creating datasets for training the C-NF model.
Using this form, we do not need to apply any further transformations to the samples during training.
This action possesses  $\phi(x)\rightarrow \phi(-x)$  symmetry, however it is  spontaneously broken at a specific parameter region.
We set $m^2=-4$ for our numerical experiments and observed spontaneous symmetry breaking in the theory by varying $\lambda$.
If we begin in the broken phase of the theory, the order parameter under consideration $\langle \phi^2\rangle$ is nonzero which approaches zero at the critical point and remains zero in the symmetric phase, as shown in \Cref{SSB}.
In HMC simulation we choose a parameter $\lambda$ and produce configurations based on the probability distribution: 
 \begin{align}
 P(\phi|\lambda)=\frac{1}{Z}e^{-S(\phi|\lambda)}\\
 \text{where,    } Z=\sum\limits_{\phi}e^{-S(\phi|\lambda)}\notag
 \end{align} 

 Each lattice configuration is a $2d$ matrix with the dimensions $L_x\times L_y$. 
For our experiment, we use $L_x=L_y=8 $ and add a periodic boundary condition to the lattice in all directions.
More information on lattice $\phi^4$ theory can be found in ref. \cite{Vierhaus2010Simulation}.
Some of the observables which we calculate on the lattice ensembles are: 
\begin{enumerate}
\item $\langle \Tilde{\phi}^2 \rangle$:
      $\Tilde{\phi}=\frac{1}{V
}\sum_x\phi(x)$

\item Correlation Function: 
\begin{align}
G_c(x)= \frac{1}{V} \sum_y[\langle\phi(y)\phi(x+y) \rangle -\langle\phi(y)\rangle \langle\phi(x+y) \rangle] \notag
\end{align}
Zero momentum Correlation Function: \\
 $C(t)=\sum_{x_1} G_c(x_1,t)$
\item Two Point Susceptibility: 
$\chi=\sum_x G(x)$

\end{enumerate}

\section{Conditional Normalizing Flow}
Normalizing flows\cite{https://doi.org/10.48550/arxiv.1505.05770} are a generative model for constructing complex distributions by transforming a simple known distribution via a series of invertible and smooth mapping $f :\mathbb{R}^d \rightarrow \mathbb{R}^d $ with inverse $
f^{-1}=g$.
If $p_z(z)$ is the prior distribution and $p_t(x)$ is the complex target distribution, then the model distribution $p_m(x;\theta)$ can be written using the change of variable formula as 
  \begin{align}
      p_m(x;\theta)=p_z(z)\big|det \frac{\partial f_{\theta}(z)^{-1}}{\partial x}\big|\\
      \text{where,  } x=f_{\theta}(z) \notag
      \label{probabilty_Trans}
  \end{align}
 Fitting a flow-based model $p_m(x;\theta)$ to a target distribution $p_t(x)$  can be accomplished by minimising their KL divergence.
The most crucial step is to build the flow so that we can calculate $\big|det \frac{\partial f_{\theta}(z)^{-1}}{\partial x}\big|$.
One such method is the affine coupling block, which divides the input $z$ into two halves and applies an affine transformation to produce upper or lower triangular Jacobians.
The transformation rules for such a building are as follows\cite{https://doi.org/10.48550/arxiv.1907.02392}: 
  \begin{align}
     x_1=z_1\odot exp(s_1(z_2)+t_1(z_2)) \notag\\
     x_2=z_2\odot exp(s_2(x_1))+t_2(x_1)
 \end{align}
where, $\odot$ represent element-wise product of two vectors.

The inverse of this coupling layer is simply computed as: 
  \begin{align}
     z_2=(x_2-t_2(x_1)) \odot exp(-s_2(x_1)) \notag\\
     z_1=(x_1-t_1(z_2)) \odot exp(-s_1(z_2))
 \end{align}
\begin{figure}[H]
      \includegraphics[width=.45\textwidth, right]{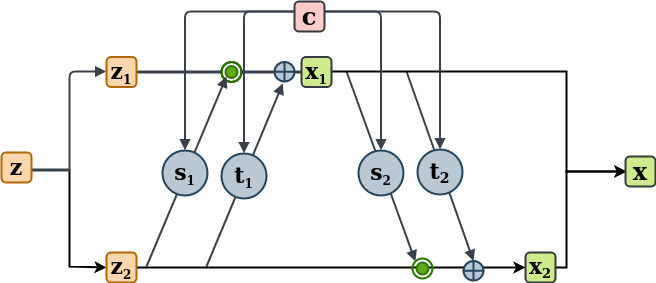}
      \caption{One affine block of Conditional Normalizing flow. Here, c is the conditional parameter. }
      \label{CNF}
\end{figure}

 Because inverting an affine coupling layer does not require the inverse of $s_1,s_2,t_1,t_2$, they can be any non-linear complex function and can thus be represented by neural networks.
Introducing a conditioning parameter in NF is not as simple as it is in GAN.
However, since $ s $ and $ t $ are only evaluated in the forward direction, we may concatenate the conditioning parameter $c$ with the input $x$ to the coupling layer in order to invert the model\cite{DBLP:journals/corr/abs-1907-02392}, as illustrated in \Cref{CNF}.
As a result, the affine coupling transformation rules become: 
    \begin{align}
     &x_1=z_1\odot exp(s_1(z_2,c)+t_1(z_2,c))\notag\\
     &x_2=z_2\odot exp(s_2(x_1,c))+t_2(x_1,c))
 \end{align}
And its inverse become
    \begin{align}
     &z_2=(x_2-t_2(x_1,c)) \odot exp(-s_2(x_1,c)) \notag\\
     &z_1=(x_1-t_1(z_2,c)) \odot exp(-s_1(z_2,c))
 \end{align}
 Let us designate the C-NF model as $f(x;c,\theta)$ and its inverse as $g(z;c,\theta)$.
The invertibility for any fixed condition $c$ is given by
 \begin{align}
     f^{-1}(.;c,\theta)=g(.;,c,\theta)
 \end{align}
 The change of variable formula become 
  \begin{align}
 p_m(x;c,\theta)=p_z(f(x;c,\theta))|det J_f(z)|^{-1}
  \end{align}
  And the loss function is the KL divergence between the model distribution $p_m(x;c,\theta)$ and target distribution $p_t(x;c)$ :
  \begin{align}
      \mathcal{L}=D_{KL}[{ p_t(x;c)||p_m(x;c,\theta)}]           
      \label{loss}
  \end{align}
  We can find the maximum likelihood network parameter  $\theta_{ML}$ using this loss. Then, for a fixed $c$, we can execute conditional generation of $x$ by sampling $z$ from $p_z(z)$  and employing the inverted network $g(z;c,\theta_{ML})$.

\section{Numerical Experiments}
This section discusses dataset preparation and the model architecture utilized in training the C-NF model. In addition, the training details of the C-NF model and the sampling process in the critical region are discussed. We train the C-NF model with different datasets and model architecture for interpolation and extrapolation. 
\subsection{Dataset}
In the non-critical region, we generate lattice configurations using HMC simulation where autocorrelation is low. In HMC simulation we use Molecular Dynamics(MD) step size=0.1 and MD trajectory length=1. For training purposes, we generate 10,000 lattice configurations for interpolation and 15,000 configurations for extrapolation of size $8\times8$ for each $\lambda$ value. 
In the critical region, for the evaluation purpose of the C-NF model, we use HMC to generate around $10^5$ lattice configurations per $\lambda$.
This serves as a baseline for comparing observables to the observable produced through the proposed sampling strategy. 

\subsection{C-NF Model Architecture}
\begin{figure}[!ht]
\centering
      \includegraphics[width=.4\textwidth]{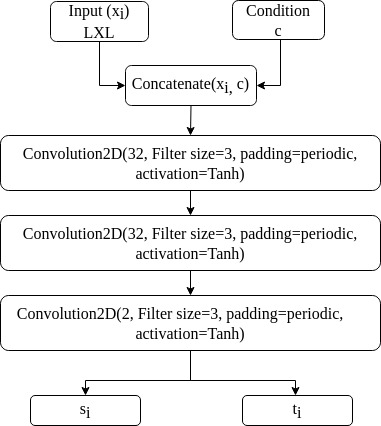}
      \caption{Architecture of the Neural Network \textbf{s} and \textbf{t} of the $i$-th affine block.}
      \label{Full_cnf}
\end{figure}

The affine coupling block displayed in \Cref{CNF} is the fundamental building component of our C-NF model.
For all neural networks $s$ and $t$, we use the same architecture.
\Cref{Full_cnf} depicts the neural networks used for $s$ and $t$. In the neural networks, we solely employ Fully Convolutional Layers. For the first two layers, we use 64 filter for interpolation and 32 for extrapolation, all of which are $3\times3$ in size. For each Convolutional layer, we employ the Tanh activation function. 
We use 8 such affine blocks while training for both interpolation and extrapolation.
For each 8 affine block, the conditional parameter $c$ is concatenated with the input to $s$ and $t$ as shown in \Cref{CNF}. 
\subsection{Training and Sampling Procedure} The loss function used for training the C-NF model is the forward KL divergence as in  \Cref{probabilty_Trans} between the model distribution $p_m(\phi;\lambda,\theta)$ and target distribution $p_t(\phi;\lambda)$:
\begin{align}
  \mathcal{L(\theta)}&=\int d\phi p_t(\phi;\lambda)(log[p_t(\phi;\lambda)]-log[p_m(\phi;\lambda,\theta)])\notag \\
  &=E_{\phi\sim p_t}(log[p_t(\phi;\lambda)]-log[p_m(\phi;\lambda,\theta)])\\
\text{where},\notag\\ &log[p_t(\phi;\lambda)]=-[S(\phi|\lambda)+Z] \notag
\end{align}

The expectation $E_{\phi\sim p_t}$ is evaluated using HMC samples from the non-critical region. 
During training, we maintain the learning rate at 0.0003 and employ the Adam optimizer.

Once training is complete, we invert the C-NF model, which we call the proposal model. The inputs to the proposal model are i) lattices from the Normal distribution $ \mathcal{N}(0,1)$ and ii)  $\lambda$ as a conditional parameter for sample generation. Outputs are the lattice configurations and probability densities for each configuration for a given $\lambda$. For the critical region we give critical $\lambda$ values as conditional parameter to the propsal model. We look at two scenarios in which a C-NF model can be either interpolated or extrapolated to the critical region. Both require different training, but the sampling technique is the same. The samples from the interpolated/extrapolated model may not exactly represent the true distribution of lattice theory $p_t(\phi|\lambda)$. As a result, we use this model as a proposal for the independent MH algorithm, which generates a Markov Chain with asymptotic convergence to the true distribution. 

\subsection{Interpolation to the Critical Region}
The $\lambda$ set used to train the C-NF model for interpolation purposes is: $\{3,3.2,3.5,3.6,3.7,3.8,5.8,6,6.5,7,8,9\}$. During training, we bypass the critical region [4.1-5.0] so that we can interpolate the model where the autocorrelation time is large for HMC simulation. 
We interpolate the trained model for multiple  $\lambda$ values belonging to the critical region  (4.1, 4.2, 4.25, 4.3, 4.35, 4.4, 4.45, 4.5, 4.55, 4.6, 4.65, 4.7, 4.8, 5.0). For each $\lambda$ values we generates one ensemble of $10^5$ configurations from the proposed method. On each ensemble we calculate different observables using bootstrap re-sampling method.
\begin{figure}[H]
    \begin{subfigure}{.5\textwidth}
        \includegraphics[
width=\textwidth]{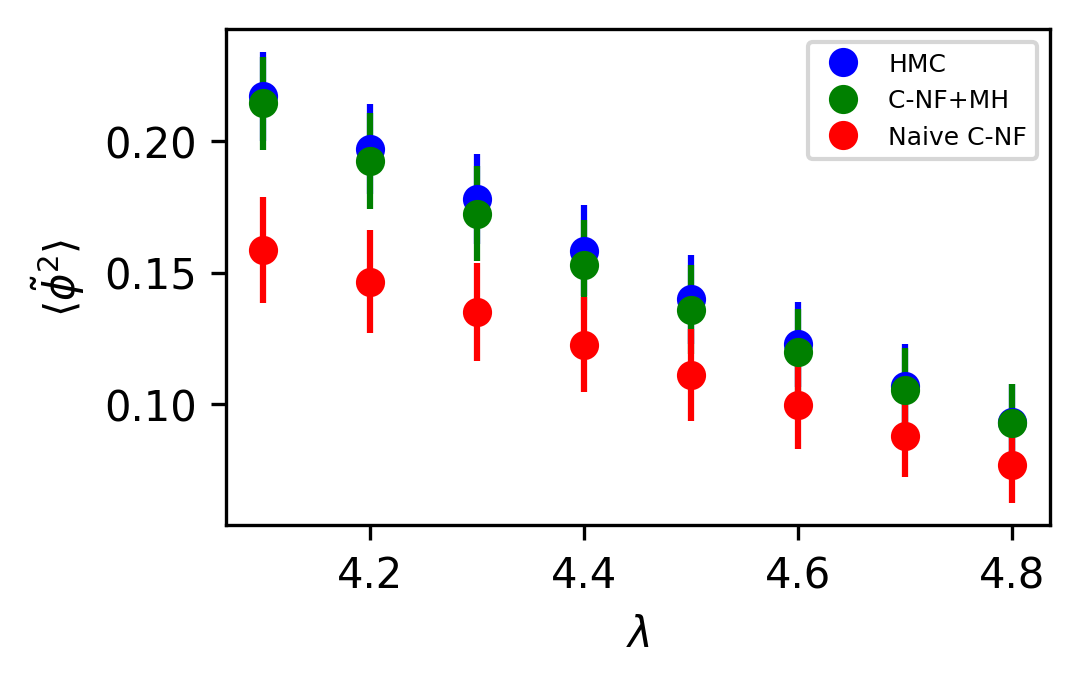}
        \caption{$\langle \Tilde{\phi}^2 \rangle$ }
        \label{msq}
    \end{subfigure}
    \begin{subfigure}{.5\textwidth}
        \includegraphics[
width=\textwidth]{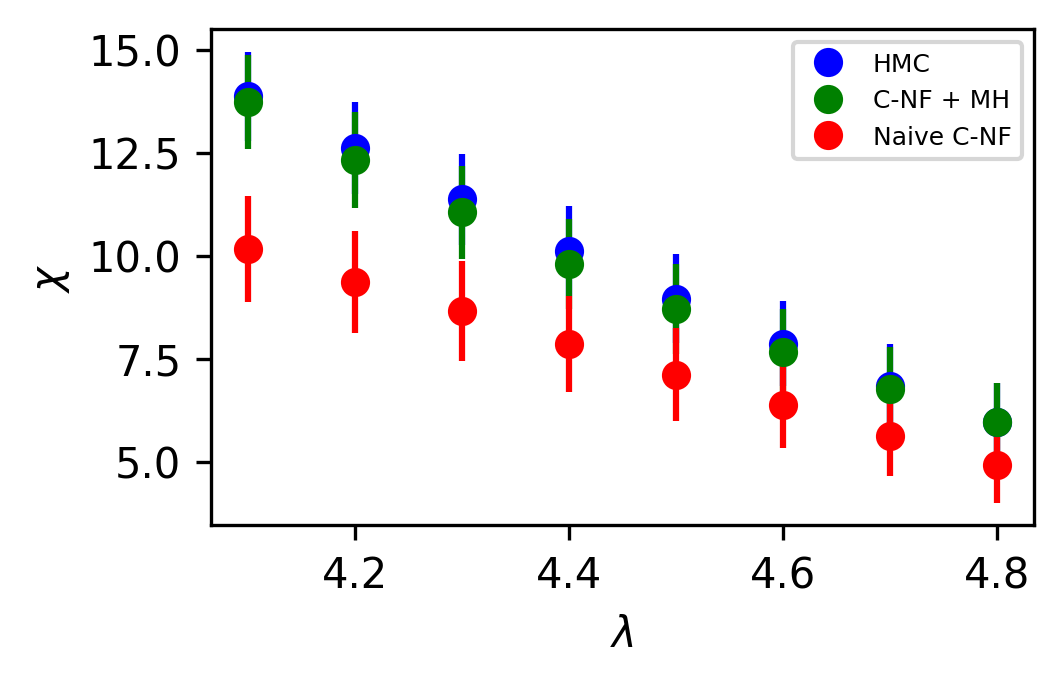}
        \caption{$\chi$}
        \label{chi}
    \end{subfigure}
  \caption{Interpolation to the critical region:-$\langle \Tilde{\phi}^2 \rangle$ and $\chi$ are calculated on samples generated from i)HMC, ii)C-NF followed by MH, and iii) Naive C-NF. The Error bars indicate standard deviation calculated using bootstrapping re-sampling with bin size 100.}
  \label{msqchi}
\end{figure}
 We compare the observables from HMC simulation and our proposed method. Observable from the Naive C-NF without MH is also shown to demonstrate the  C-NF model's proximity to the true distribution. In \Cref{msqchi} we plot two observables  $\langle \Tilde{\phi}^2 \rangle$ and $\chi$ for the interpolated $\lambda$ values. Although the naive C-NF model has biases, MH can eliminate them, and both observables match pretty well within the statistical uncertainty.
In the Appendix, we present a table of numerical values of the observables with errors. 
 \begin{figure*}[!ht]
    \begin{subfigure}{.5\textwidth}
      \includegraphics[
width=\textwidth]{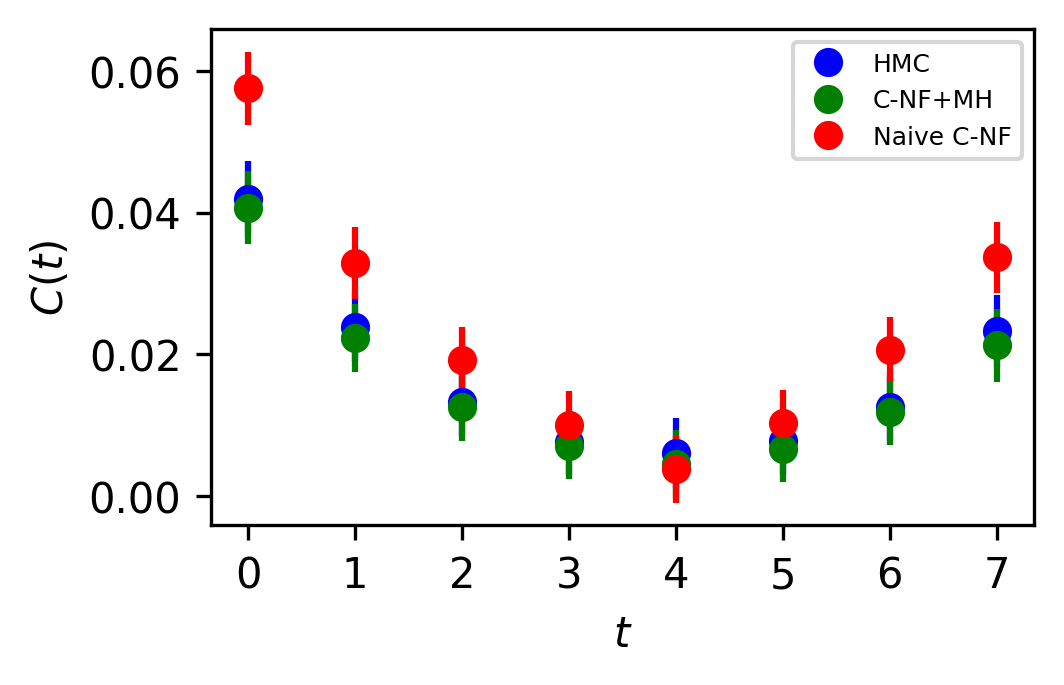}
      \caption{$\lambda=4.3$}
      \label{fig:f1}
    \end{subfigure}
    \begin{subfigure}{.5\textwidth}
        \includegraphics[
width=\textwidth]{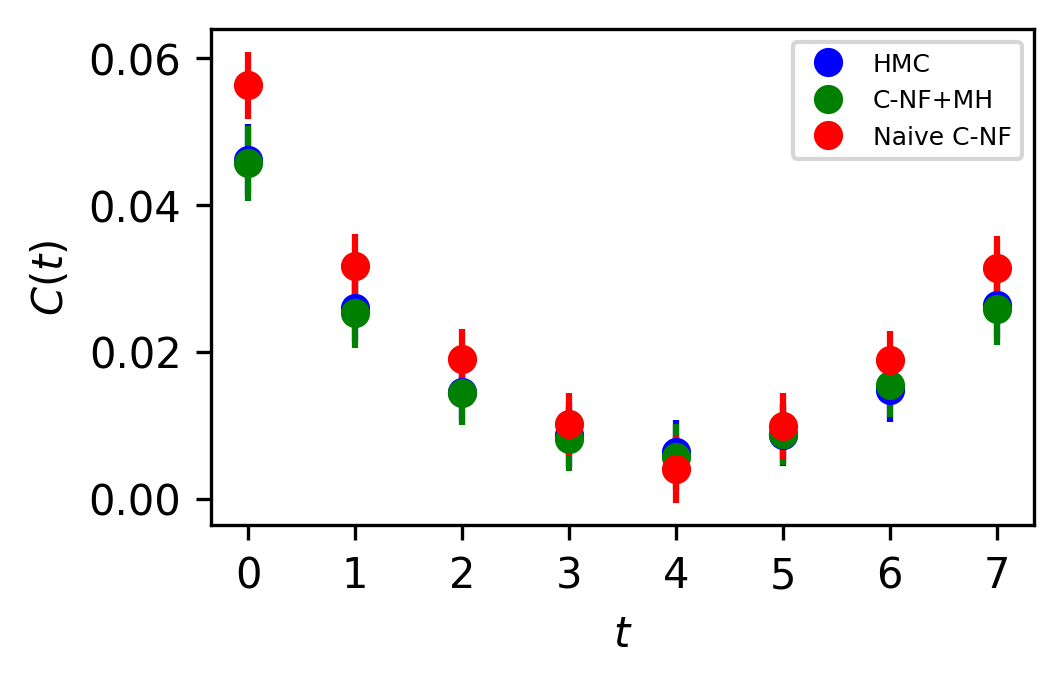}
        \caption{$\lambda=4.5$}
        \label{fig:f2}
    \end{subfigure}

    \begin{subfigure}{.5\textwidth}
        \includegraphics[
width=\textwidth]{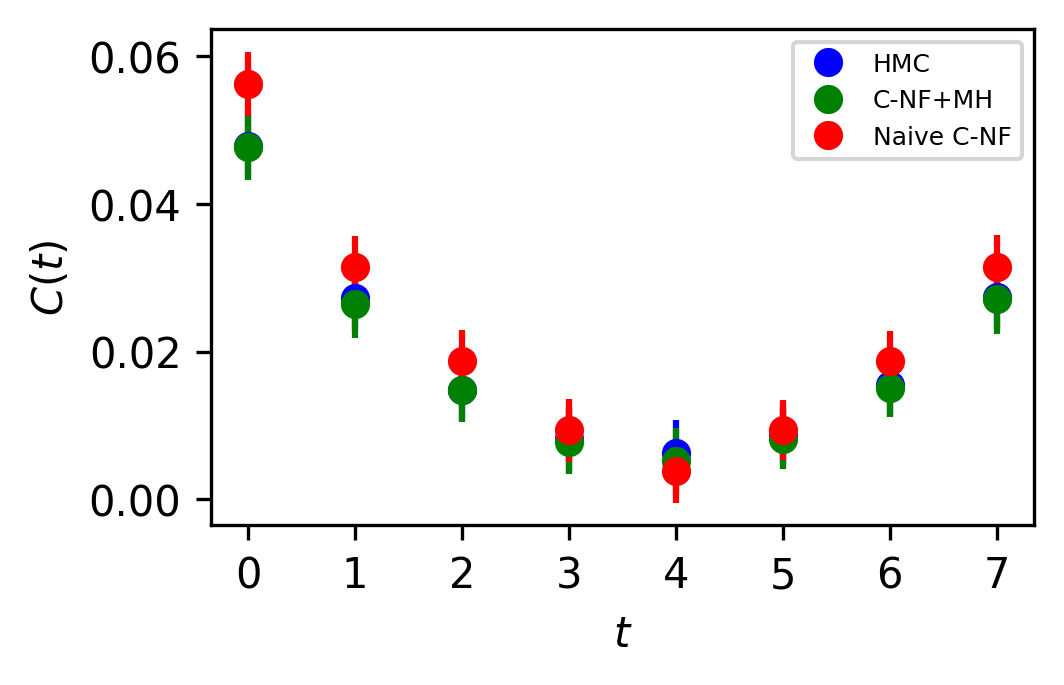}
        \caption{$\lambda=4.55$}
        \label{fig:f3}
    \end{subfigure}
     \begin{subfigure}{.5\textwidth}
        \includegraphics[
width=\textwidth]{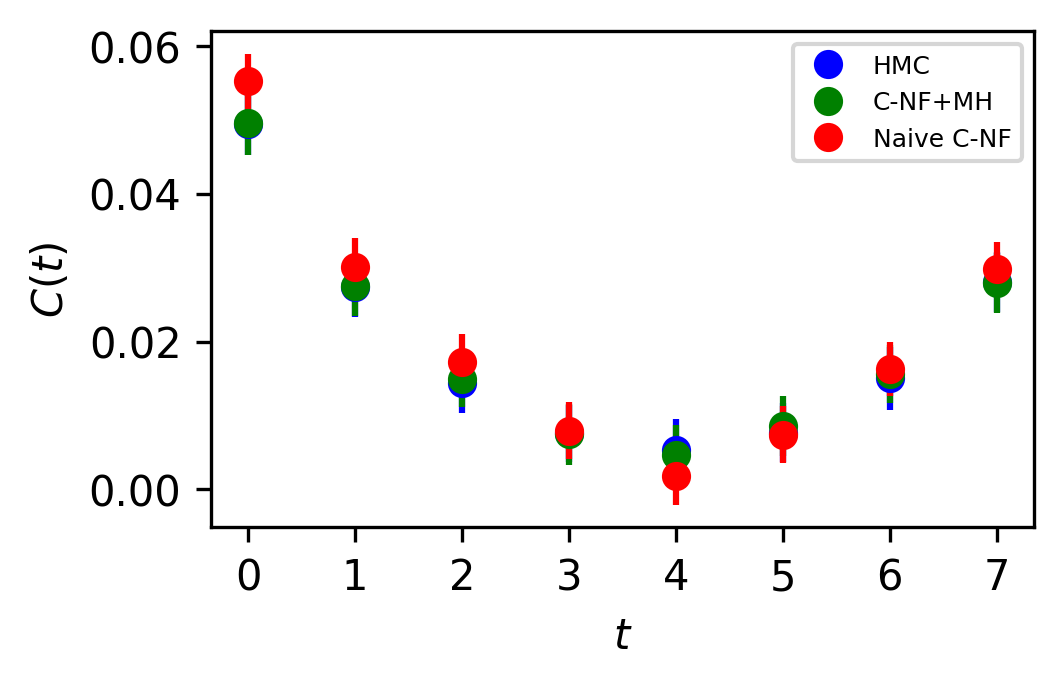}
        \caption{$\lambda=4.70$}
        \label{fig:f4}
    \end{subfigure}
\caption{Interpolation-:Zero momentum Correlation function calculated on samples generated from i)HMC, ii)C-NF followed by MH, and iii) Naive C-NF. The Error bars indicate $95\%$ confidence interval calculated using bootstrapping re-sampling method with bin-size 100.}
\label{intp_corr}
\end{figure*}
In \Cref{intp_corr}, the two point zero momentum correlation function ${C(t)}$ is shown for four different $\lambda$ values. For these $\lambda$ values, we can observe that the correlation function is non-vanishing, which is a unique property of the critical region. The plots for other critical $\lambda$ values are included in the Appendix.

In \Cref{hist} we have also shown the histogram of $\Tilde{\phi}$ for a particular critical $\lambda=4.6$. This demonstrates that MH can eliminate this kind artefacts created by the C-NF model. 

We displayed plots of various observables on both phases around the critical point for the interpolation. We found that the observables estimated using our technique and the HMC match quite well.
\begin{figure}[!ht]
    \begin{subfigure}{.5\textwidth}
      \includegraphics[width=\textwidth]{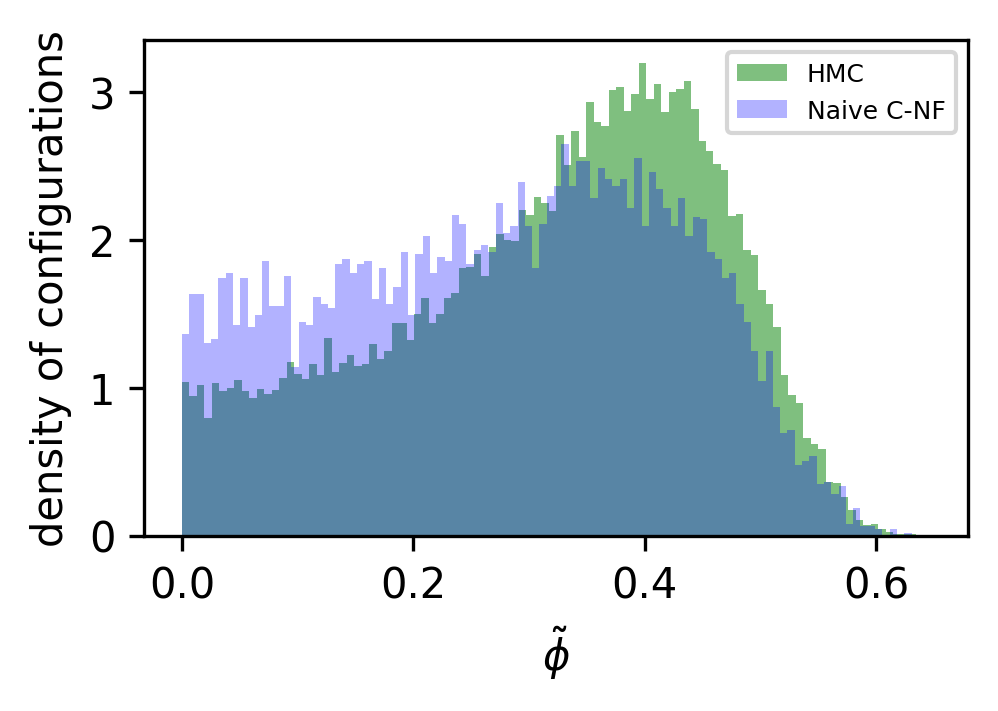}
      \label{fig:f5}
    \end{subfigure}
    \begin{subfigure}{.5\textwidth}
      \includegraphics[width=\textwidth]{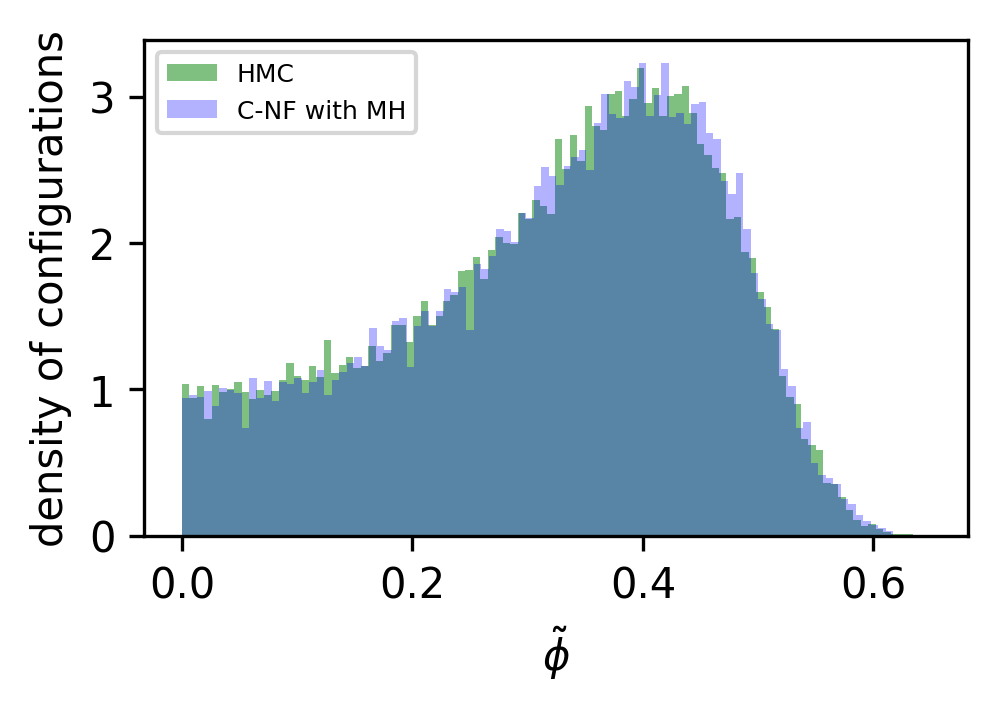}
      \label{fig:f6}
    \end{subfigure}
\caption{Histogram of $\Tilde{\phi}$ for $\lambda=4.6$ from the a)naive C-NF model and b) C-NF with MH is compared against HMC results with a bin-size=100.}
\label{hist}
\end{figure}

\subsection{Extrapolation to the Critical Region}
\begin{figure}[!ht]
    \begin{subfigure}{.5\textwidth}
        \includegraphics[
width=\textwidth]{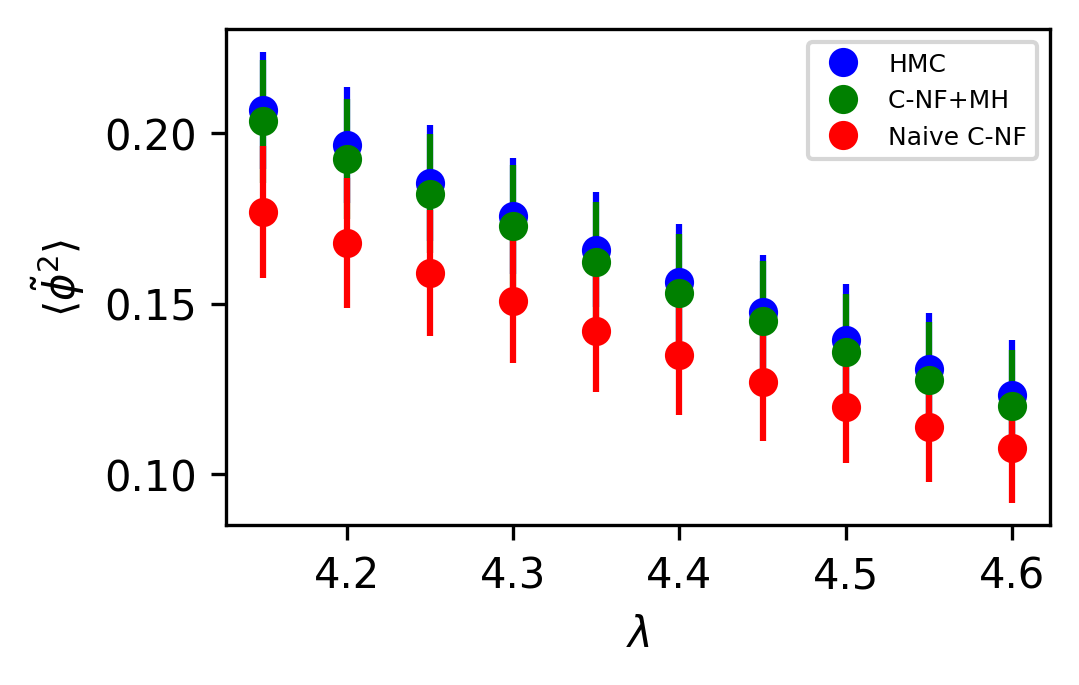}
        \caption{$\langle \Tilde{\phi}^2 \rangle$ }
        \label{fig:f7}
    \end{subfigure}
    \begin{subfigure}{.5\textwidth}
        \includegraphics[
width=\textwidth]{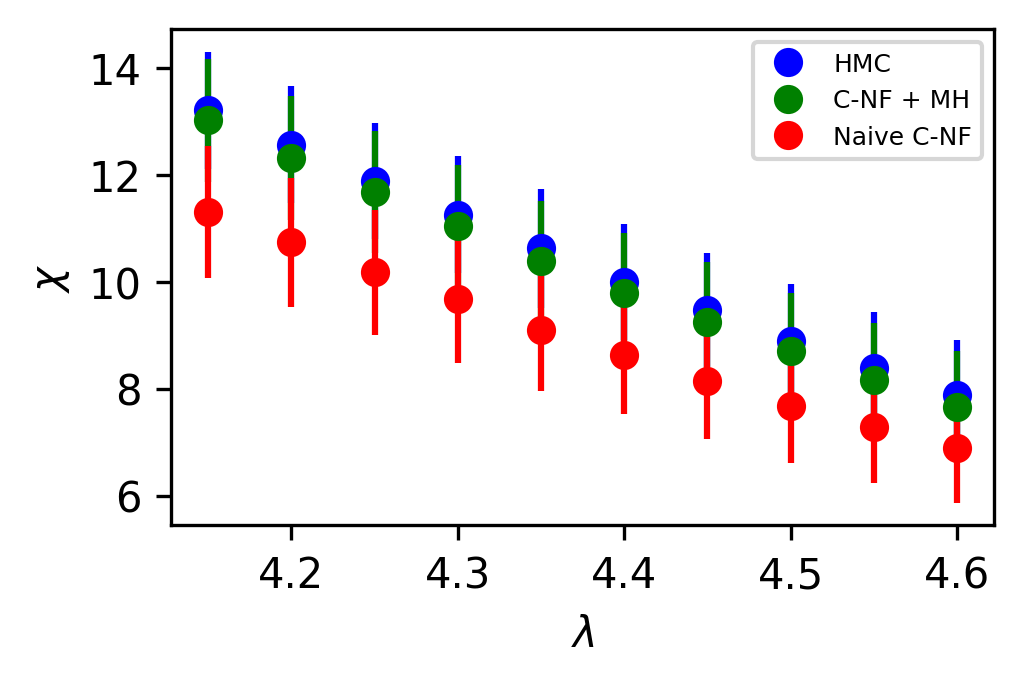}
        \caption{$\chi$}
        \label{fig:2nd}
    \end{subfigure}
  \caption{ Extrapolation to the critical region-: $\langle \Tilde{\phi}^2 \rangle$ and $\chi$ are calculated on samples generated from i)HMC, ii)C-NF followed by MH, and iii) Naive C-NF. The Error bars indicate standard deviation calculated using bootstrapping re-sampling with bin size 100.}
  \label{extra_chi}
\end{figure}

\begin{figure*}[!ht]
    \begin{subfigure}{.5\textwidth}
      \includegraphics[width=\textwidth]{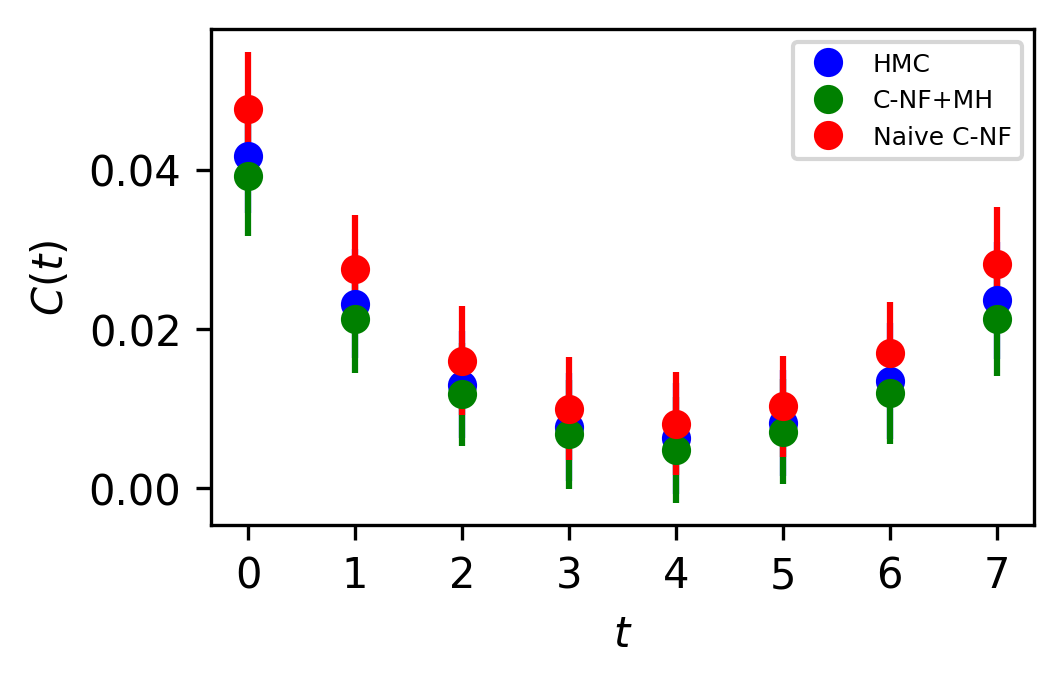}
      \caption{$\lambda=4.20$}
      \label{fig:9}
    \end{subfigure}
    \begin{subfigure}{.5\textwidth}
      \includegraphics[width=\textwidth]{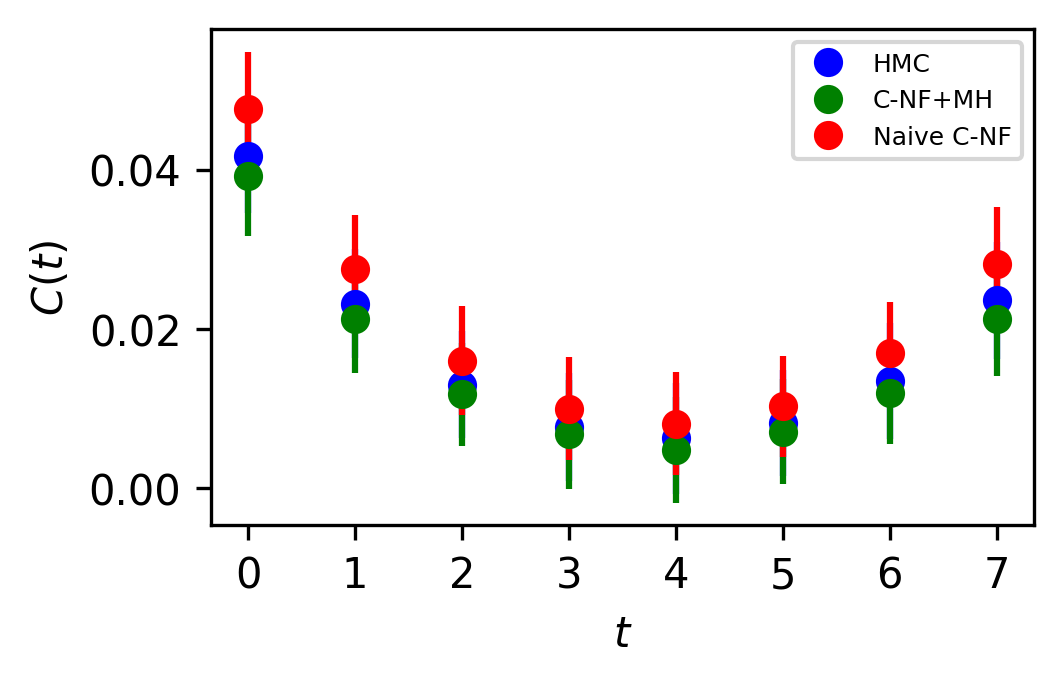}
      \caption{$\lambda=4.30$}
      \label{fig:f10}
    \end{subfigure}
    \begin{subfigure}{.5\textwidth}
        \includegraphics[
width=\textwidth]{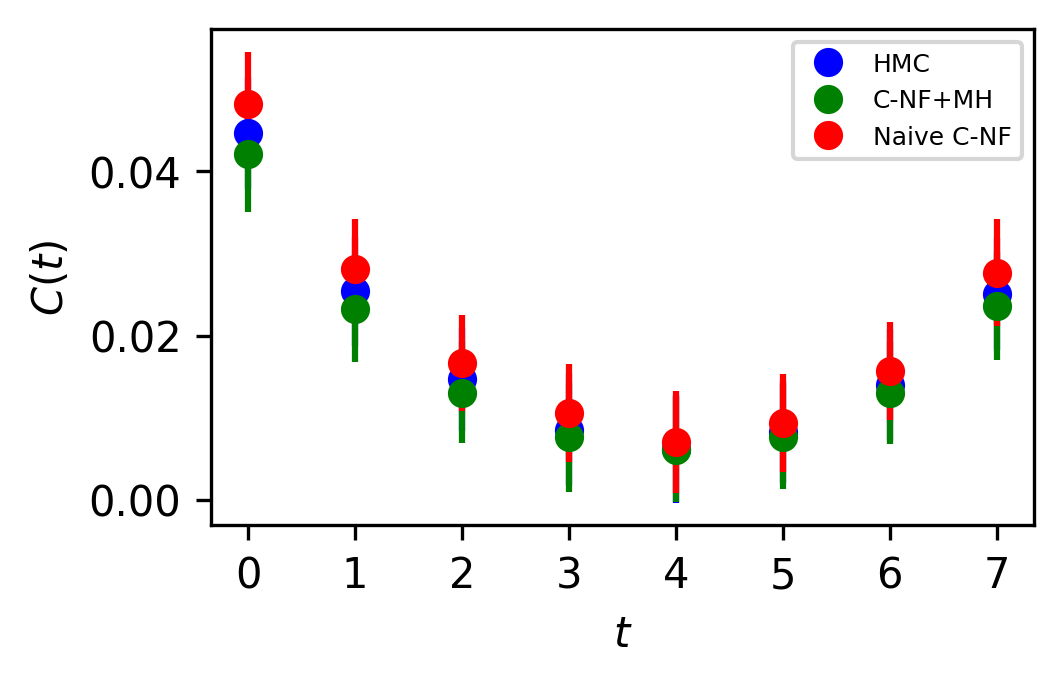}
        \caption{$\lambda=4.40$}
        \label{fig:f11t}
    \end{subfigure}
    \begin{subfigure}{.5\textwidth}
        \includegraphics[
width=\textwidth]{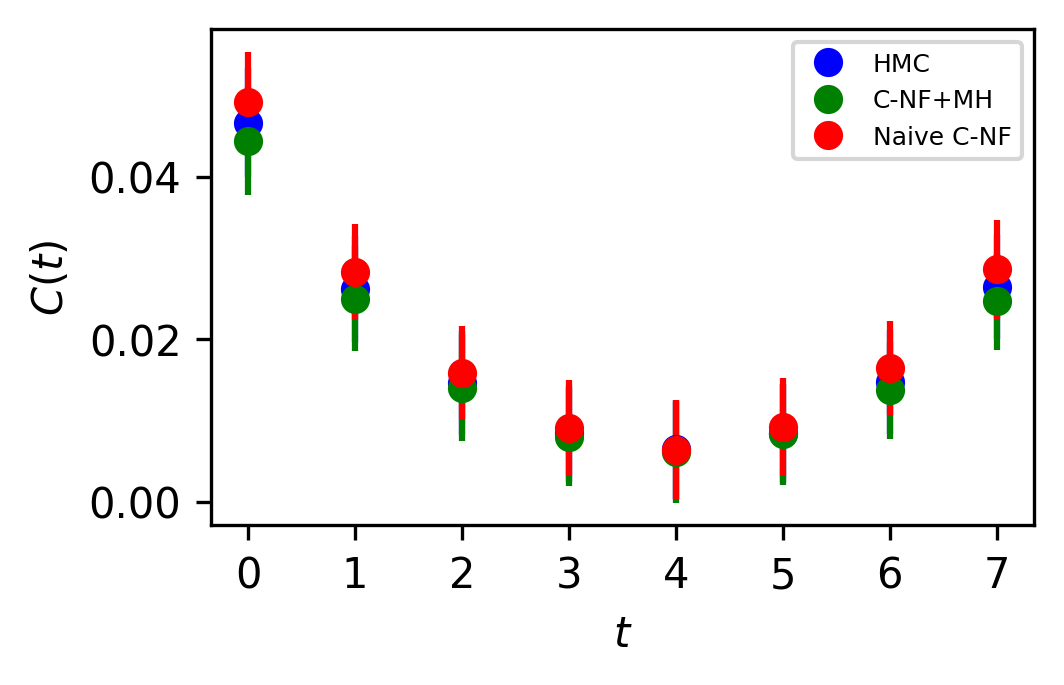}
        \caption{$\lambda=4.45$}
        \label{fig:f12}
    \end{subfigure}
\caption{Extrapolation:-Zero momentum Correlation function calculated on samples generated from i)HMC, ii)C-NF followed by MH, and iii) Naive C-NF. The Error bars indicate $95\%$ confidence interval calculated using bootstrapping re-sampling method with bin-size 100.}
\label{corr_ext}
\end{figure*}

The $\lambda$ set used to train the C-NF model for the extrapolation is :$\{3,3.1,3.2,3.3,3.4,3.5,3.6,3.7,3.8,3.9\}$. After training the model in the broken phase, we extrapolate it in the critical region around $\lambda=4.6$, which we take to be the critical region's midpoint. The model is extrapolated for five distinct $\lambda$ values:[4.2,4.3,4.4,4.5,4.6]. For each $\lambda$ values we generates one ensemble of $10^5$ configurations from our method. Again, we calculate several observables for each ensemble using the bootstrap re-sampling method and compare them to the HMC results. 
We find that the size of the training dataset needs to be increased for extrapolation in order to achieve a C-NF model that is close to the true distribution in the broken phase. We plot the observables  $\langle \Tilde{\phi}^2 \rangle$ and $\chi$  in \Cref{extra_chi}  and the zero momentum  correlation function is plotted in \Cref{corr_ext} for four different $\lambda$ in the critical region .

\begin{figure}[H]
\centering
      \includegraphics[width=.5\textwidth]{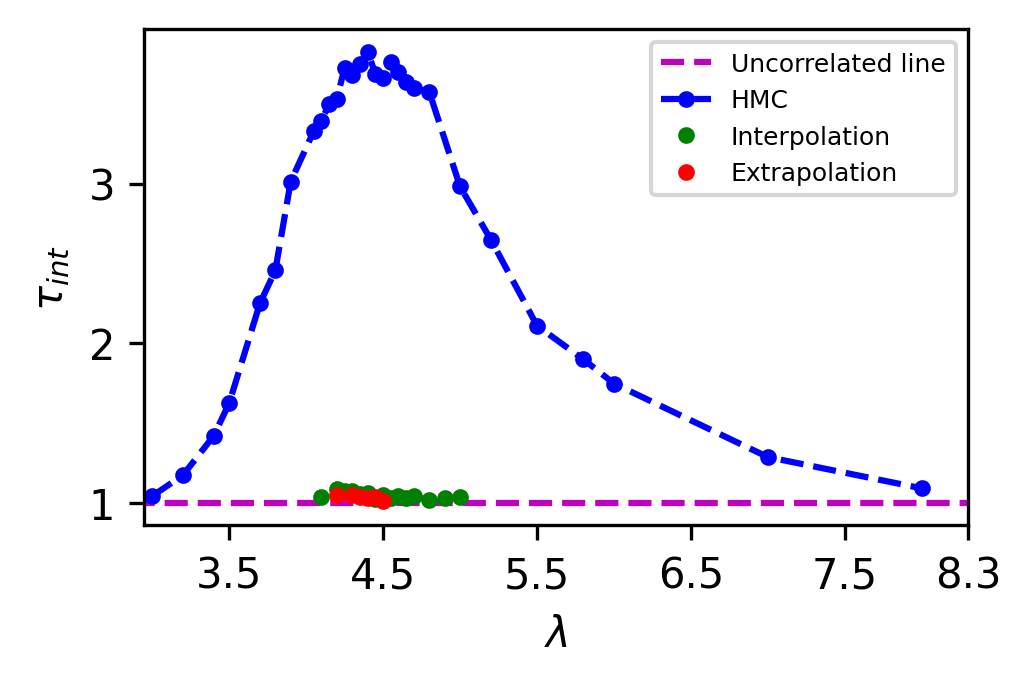}
      \caption{Integrated Auto-correlation time calculated on samples generated from HMC and C-NF model with MH. The straight line represents the autocorrelation time for uncorrelated samples.}
      \label{autocorr}
\end{figure}

The naive C-NF model produces configurations that are inherently uncorrelated. Since we obtain a Markov Chain after applying MH, therefore we can't guarantee the same. With a $25-40\%$ acceptance rate from the model, we see negligible correlation between the samples. The integrated autocorrelation time for $\chi$ from our proposed approach and HMC is plotted in \Cref{autocorr}. It shows that we successfully reduce the Autocorrelation time for the Markov chain.The comparison is not absolute because HMC is affected by algorithmic settings. But we want to show that for the critical region, there is almost no correlation between samples in the  Markov Chains obtained from both interpolation and extrapolation.

\section{Cost Analysis}The sampling algorithm for the baseline approach (HMC) and the suggested method is vastly different; thus, a direct cost comparison is opaque. Nonetheless, we separate the simulation cost for the proposed technique into two components: Training time and sample generation time.
On a Colab Tesla P100 GPU, the training time for the C-NF model for interpolation or extrapolation is roughly 5-6 hours.
However, sample generation is very fast for the C-NF model.
Generating one Markov Chain of $10^5$ configuration takes 5-7 minutes with a 25-40$\%$ acceptance rate. Due to the short generation time, a low acceptance rate is acceptable until the autocorrelation time increases.

Once the C-NF model has been trained, it can be employed repeatedly to generate configurations for a wide range of $\lambda$ values.
From a single training of the C-NF(interpolated) model, we have generated configurations for 13 $\lambda$'s in the critical region. So, our approach outperforms HMC for sampling at multiple $\lambda$ values in the critical region.

\section{Conclusion}
The critical slowing down problem prevents generating a large ensemble in the critical region of a lattice theory. In order to resolve this, we employ a Conditional normalizing flow trained on HMC samples with low autocorrelation and generate samples in the critical region. In order to learn a general distribution over parameter $\lambda$, we train the C-NF model away from the critical point of lattice $\phi^4$ theory. This model is interpolated in the critical region and serves as a proposal for the MH algorithm to generate a Markov chain. The degree to which the extrapolated or interpolated model resembles the actual distribution determines the acceptance rate in the critical region. In order to achieve a high acceptance rate and prevent the development of autocorrelation, the C-NF model must be trained adequately. The C-NF model generates uncorrelated samples, and we trained well enough to get 25-45$\%$ acceptance rate. With this much acceptance rate, we found no correlation between configuration in the Markov chain. 
As a result, our method significantly mitigates the critical slowing down problem. Aside from that, our method can be highly efficient when we need interpolation/extrapolation to numerous  $\lambda$ values in the critical region.
Since lattice gauge theory requires extrapolation to the critical region, we have likewise extrapolated $\phi^4$ lattice theory to the critical region.
We observe high agreement between observables estimated using the suggested technique and HMC simulation for both interpolation and extrapolation. 
\bibliography{ NFbib}
\onecolumn
\section{ Appendix}

\fontsize{8}{12}\selectfont
\begin{table}[H]
\begin{tabularx}{1.0\textwidth}{|X|X|X|X|X|X|X|}
 \hline
 \multirow{2}{4em}{$\lambda$} &  \multicolumn{3}{c|}{$\langle \Tilde{\phi}^2 \rangle$} & \multicolumn{3}{c|}{$\chi_2$  } \\

\cline{2-7}
 & HMC    &  C-NF with MH & Naive C-NF  & HMC  &  C-NF with MH   & Naive C-NF  \\
 \hline
 4.10 & $ 0.2147 \pm 0.0179 $ &$ 0.2169 \pm 0.0169$&$ 0.1586 \pm 0.0205$&$ 13.7446 +\pm 0.8059$&  $13.7446 \pm 0.8059$& $10.1557 \pm 0.9181$ \\
 \hline 
 4.20 & $0.1925 \pm 0.0181$ &$0.1970 \pm 0.0173$& $ 0.1463 \pm 0.0199$& $12.3272 \pm 0.8103$ & $12.6235 \pm 0.7822$&$9.3694 \pm 0.8844$\\
 \hline
 4.25 & $0.1819 \pm 0.0090$&$ 0.1872 \pm 0.0087 $& $ 0.1411 \pm 0.0097$&$11.6485 \pm 0.5747$&$11.9992 \pm 0.5566$& $9.0395 \pm 0.6191$\\
 \hline
 4.30 &$ 0.1724 \pm 0.0181$ &$ 0.1781 \pm 0.0176$& $ 0.1348 \pm 0.0190$ 
 &$11.0331 \pm 0.8188$&$11.3775 \pm 0.7762$&$8.6260 \pm 0.8674$\\
 \hline
 4.35 & $ 0.1622 \pm 0.0177$&$0.1681 \pm 0.0174$&$ 0.1289 \pm 0.0184$&$10.3827 \pm 0.5648$&$10.7440 \pm 0.5640$&$8.2559 \pm 0.5894$\\
 \hline
 4.40 & $0.1530 \pm 0.0178$ &$ 0.1583 \pm 0.0172$&$0.1225 \pm 0.0182$&$9.7920 \pm 0.5529$&$10.1284 \pm 0.5541$&$7.8501 \pm 0.5867$\\
 \hline
 4.45 &$ 0.1446 \pm 0.0173$&$0.1490 \pm 0.0170$&$0.1166 \pm 0.0176$&$9.2557 \pm 0.5440$&$ 9.5575 \pm 0.5516$&$7.4716 \pm 0.5637$\\
 \hline
 4.50 &$ 0.1357 \pm 0.0172$&$ 0.1399 \pm 0.0170$&$ 0.1115 \pm 0.0175$&$ 8.7005 \pm 0.5446$&$8.9528 \pm 0.5442$&$7.1308 \pm 0.5479$\\
 \hline
 4.60& $0.1197 \pm 0.0163$&$0.1227 \pm 0.0163$&$0.0996 \pm 0.0165$&$ 7.6629 \pm 0.5272$&$ 7.8601 \pm 0.5188$&$6.3753 \pm 0.5271$\\
 \hline
 4.65 & $ 0.1127 \pm 0.0157$&$0.1151 \pm 0.0162$&$0.0939 \pm 0.0159$&$7.1991 \pm 0.5041$&$7.3486 \pm 0.5126$&$ 6.0062 \pm 0.5109$\\
 \hline
 4.70 &$ 0.1054 \pm 0.0153$&$ 0.1069 \pm 0.0157$&$0.0879 \pm 0.0154$&$ 6.7575 \pm 0.4978$&$ 6.8422 \pm 0.5075$&$5.6177 \pm 0.4966$\\
 
 \hline
 4.8 & $0.0931 \pm 0.0148$ & $0.0931 \pm 0.0148$ & $0.0769 \pm 0.0140$ & $5.9530 \pm 0.4717$ & $ 5.9569 \pm 0.4671$& $4.9223 \pm 0.4590$\\
\hline
5.0 &  $ 0.0733 \pm 0.0128$& $0.0704 \pm 0.0128$ & $0.0593 \pm 0.0121$ & $4.6935  \pm 0.4131$ & $4.4963 \pm 0.4180$ & $3.7920  \pm 0.3886$\\
\hline
\end{tabularx}
\caption{\label{tab:interpolation} Two observables calculated on samples  generated from i) HMC ii) C-NF with MH and iii) Naive C-NF .The Error indicates standard deviation calculated using bootstrapping re-sampling with bin-size 100.}
\end{table}

\fontsize{8}{12}\selectfont
\begin{table}[H]
\begin{tabularx}{1.0\textwidth}{X|X|X|X|X|X|X|X|X|X|X}
\hline
 $\lambda$ & 4.1 & 4.2 & 4.3 & 4.4 & 4.5 & 4.6 & 4.7 & 4.8 & 4.9 & 5.0\\
  \hline
 $\tau_{int}$ &1.033 & 1.084 & 1.071 & 1.059 & 1.048 & 1.039 & 1.041 & 1.018 & 1.030 & 1.037\\
 \hline
 \end{tabularx}
 \caption{\label{tab:interpolation_int} Integrated Autocorrelation time for $\langle \Tilde{\phi}^2 \rangle$ from the interpolated C-NF model after applying MH. }
 \end{table}
 \fontsize{8}{12}\selectfont
\begin{table}[H]
\begin{tabularx}{1.0\textwidth}{X|X|X|X|X|X}
\hline
 $\lambda$  & 4.2 & 4.3 & 4.4 & 4.5 & 4.6 \\
  \hline
 $\tau_{int}$ & 1.046& 1.049 & 1.027 & 1.032 & 1.010 \\
 \hline
 \end{tabularx}
 \caption{\label{tab:interpolation_extra} Integrated Autocorrelation time for $\langle \Tilde{\phi}^2 \rangle$ from the extrapolated C-NF model after applying MH. }
 \end{table}
\fontsize{8}{12}\selectfont
\begin{table}[H]
\begin{tabularx}{1.0\textwidth}{|X|X|X|X|X|X|X|}
 \hline
 \multirow{2}{4em}{$\lambda$} &  \multicolumn{3}{c|}{$\langle \Tilde{\phi}^2 \rangle$} & \multicolumn{3}{c|}{$\chi_2$  } \\

\cline{2-7}
 & HMC    &  C-NF with MH & Naive C-NF  & HMC  &  C-NF with MH   & Naive C-NF   \\
 \hline
 4.20 & $0.1925 \pm 0.0181$ & $ 0.1966 \pm 0.0170$&$0.1676 \pm 0.0189$& $12.3272 \pm 0.8103$&$12.5670 \pm 0.5490$ & $10.7348 \pm 0.5989$\\
 \hline
 4.30 &$ 0.1724 \pm 0.0181$ & $  0.1756 \pm 0.0171$ &$ 0.1509 \pm 0.0177$
 &$11.0331 \pm 0.8188$&$11.2388 \pm 0.5571$&$9.6593 \pm 0.5868$\\
 \hline

 4.40 & $0.1530 \pm 0.0178$ &$0.1562 \pm 0.0171$&$ 0.1348 \pm 0.0173$&$9.7920 \pm 0.5529$&$9.9943 \pm 0.5506$&$8.6316 \pm 0.5510$\\
 \hline

 4.50 &$ 0.1357 \pm 0.0172$&$ 0.1393 \pm 0.0167$&$ 0.1200 \pm 0.0164$&$ 8.7005 \pm 0.5446$&$8.9094 \pm 0.5293$&$7.6643 \pm 0.5342$\\
 \hline
 4.60& $0.1197 \pm 0.0163$&$0.1233 \pm 0.0157$&$0.1075 \pm 0.0158$&$ 7.6629 \pm 0.5272$&$7.8881 \pm 0.5143$&$ 6.8783 \pm 0.4992$\\
 \hline

\end{tabularx}

\caption{\label{tab:extrapolation} Two observables calculated on samples  generated from i)HMC ii)C-NF with MH and iii) Naive C-NF .The Error bars indicates standard deviation calculated using bootstrapping re-sampling with binsize 100. }
\end{table}

 \begin{figure*}[!ht]
    \begin{subfigure}{.5\textwidth}
      \includegraphics[width=\textwidth]{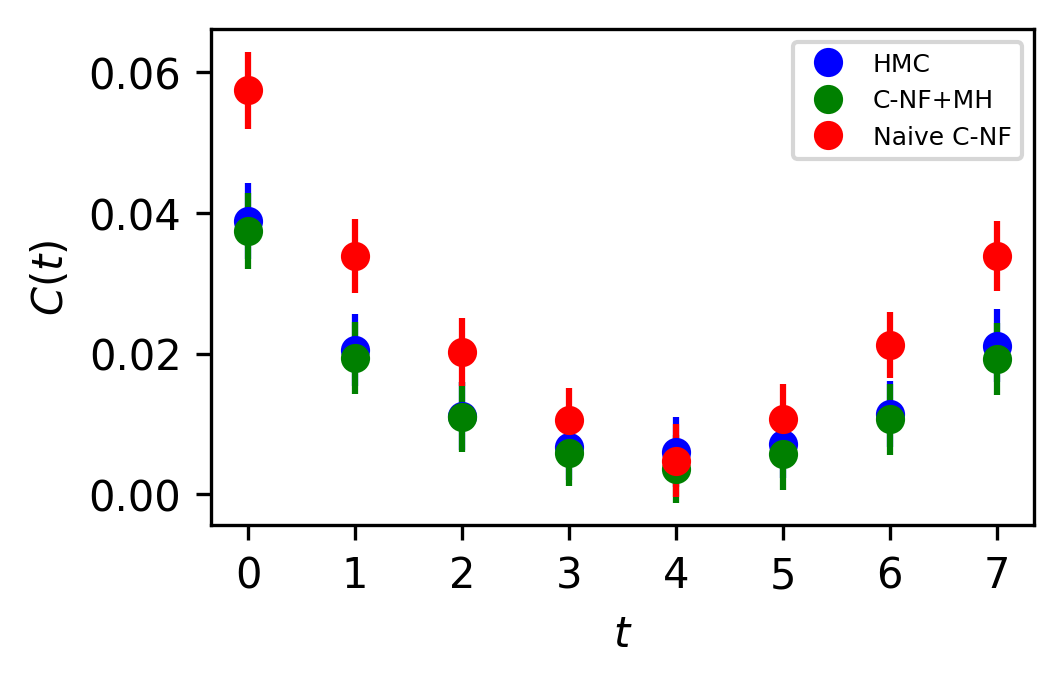}
      \caption{$\lambda=4.2$}
    \end{subfigure}
    \begin{subfigure}{.5\textwidth}
      \includegraphics[width=\textwidth]{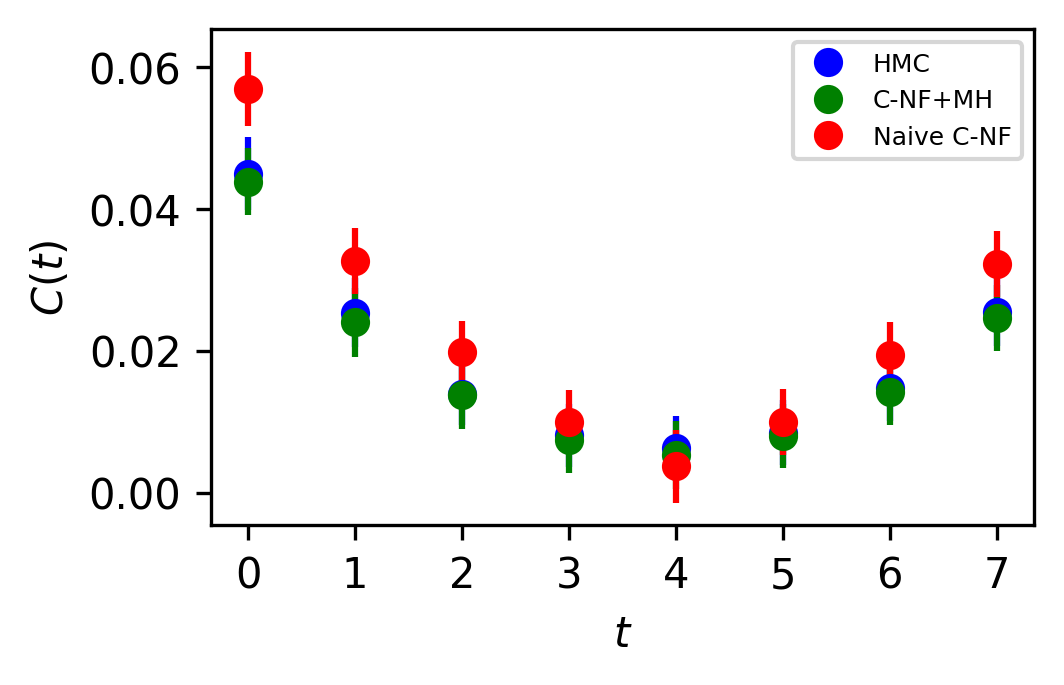}
      \caption{$\lambda=4.4$}
    \end{subfigure}
    \begin{subfigure}{.5\textwidth}
      \includegraphics[width=\textwidth]{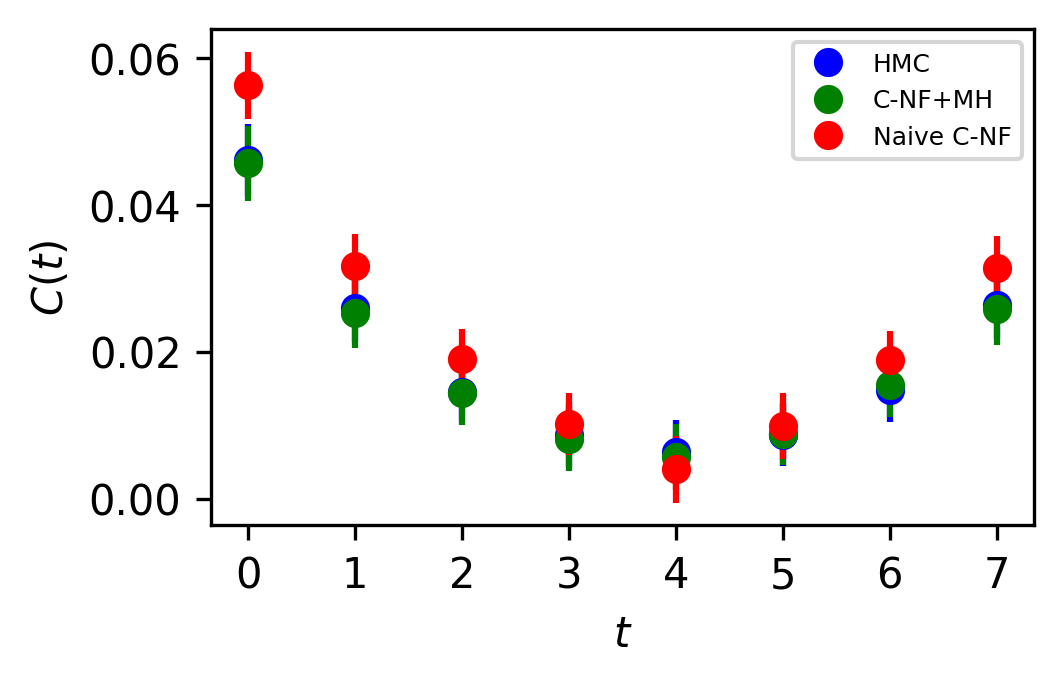}
      \caption{$\lambda=4.5$}
    \end{subfigure}
    \begin{subfigure}{.5\textwidth}
      \includegraphics[width=\textwidth]{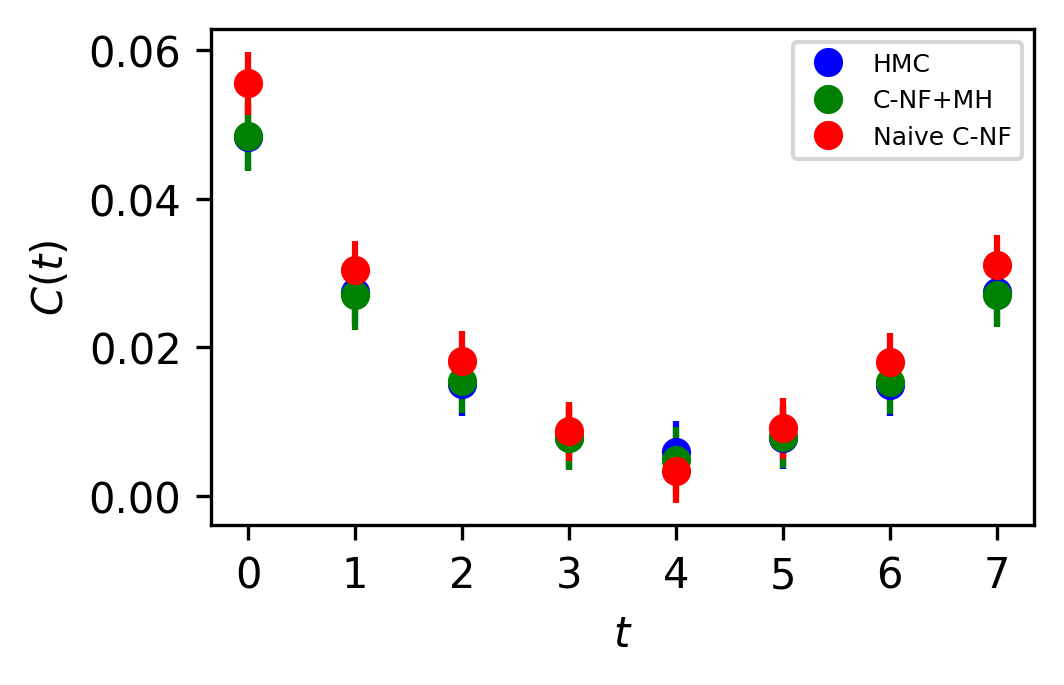}
      \caption{$\lambda=4.6$}
    \end{subfigure}
        \begin{subfigure}{.5\textwidth}
      \includegraphics[width=\textwidth]{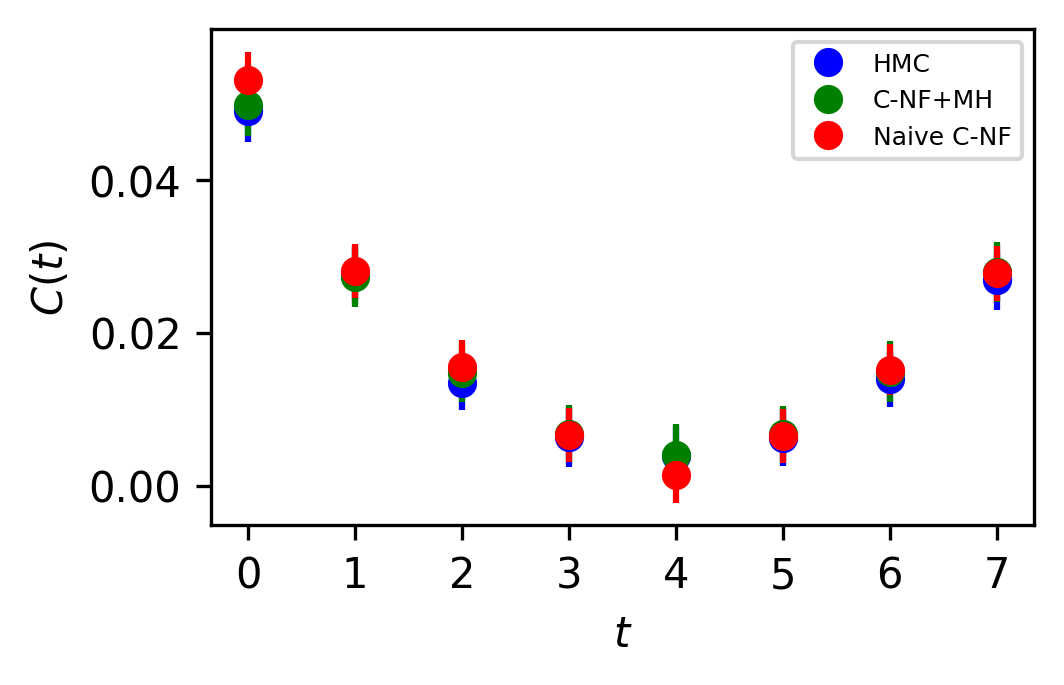}
      \caption{$\lambda=4.8$}
    \end{subfigure}
        \begin{subfigure}{.5\textwidth}
      \includegraphics[width=\textwidth]{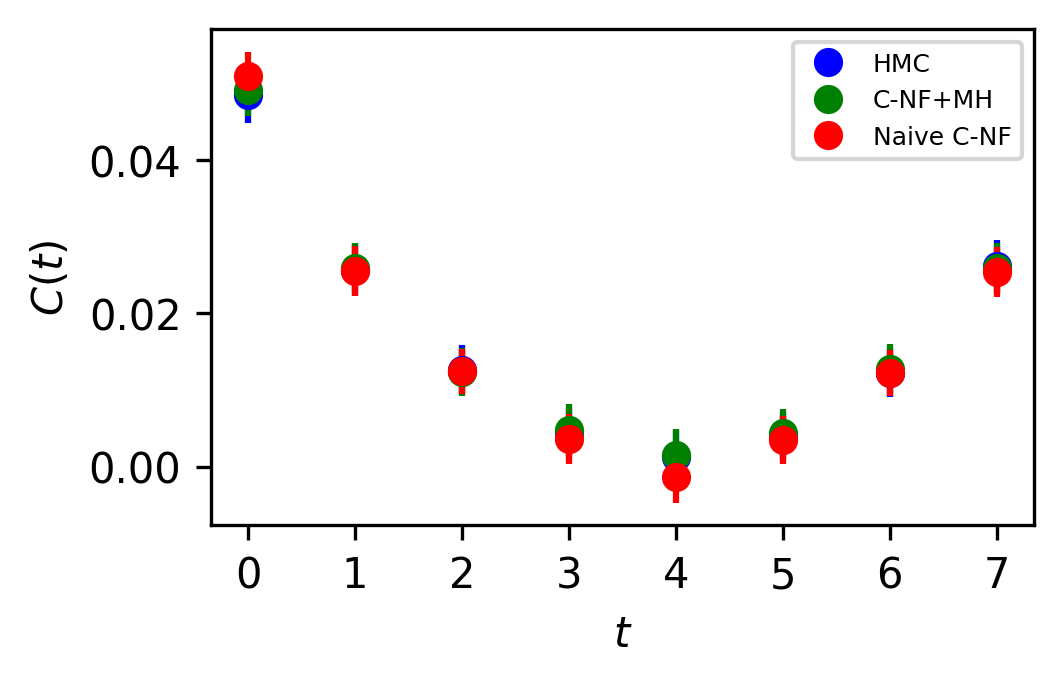}
      \caption{$\lambda=5.0$}
    \end{subfigure}
\caption{Interpolation-:Zero momentum Correlation function calculated on samples generated from i)HMC, ii)C-NF followed by MH, and iii) Naive C-NF. The Error bars indicate $95\%$ confidence interval calculated using bootstrapping re-sampling method with bin-size 100.}
\label{intp_corr_appndix}
\end{figure*}

\end{document}